\documentclass[twocolumn,superscriptaddress,showpacs,amsmath,amssymb,prb]{revtex4}
\usepackage{epsfig}
\usepackage{graphics}
\usepackage{color}
\begin{document}
\title{Resistive transition in granular disordered high-T$_c$ superconductors: \\a numerical study}
\author{L. Ponta}
\email{linda.ponta@polito.it}
\affiliation{Physics Department,  Politecnico di Torino, \\Corso Duca degli Abruzzi 24, 10129 Torino, Italy}
\author{A. Carbone}
\email{anna.carbone@polito.it}
\affiliation{Physics Department,  Politecnico di Torino, \\Corso Duca degli Abruzzi 24, 10129 Torino, Italy}
\author{M. Gilli}
\email{marco.gilli@polito.it}
\affiliation{Electronic Engineering Department, Politecnico di Torino, \\Corso Duca degli Abruzzi 24, 10129 Torino, Italy}
\author{P. Mazzetti}
\email{piero.mazzetti@polito.it}
\affiliation{Physics Department, Politecnico di Torino, \\Corso Duca degli Abruzzi 24, 10129 Torino, Italy}

\date{\today}

\begin{abstract}
The resistive transition of granular high-T$_c$  superconductors, characterized by either weak  (YBCO-like) or  strong  (MgB$_2$-like) links, occurs through a series of avalanche-type current density rearrangements. These rearrangements correspond to the creation  of resistive layers, crossing the whole specimen approximately orthogonal to the current density direction, due to the simultaneous transition of a large number of weak-links or grains. The present work
shows that exact solution of the Kirchhoff equations for strongly and weakly linked networks of nonlinear resistors, with Josephson junction characteristics, yield the subsequent formation of resistive layers within the superconductive matrix as temperature increases.
Furthermore, the voltage noise observed at the transition is related to the resistive layer formation process. The noise intensity is estimated from the superposition of voltage drop elementary events related to the subsequent resistive layers. At the end of the transition, the layers mix-up,  the step amplitude decreases and  the resistance curve smoothes. This results in the suppression of noise, as experimentally found. Remarkably, a  scaling law for the noise intensity with the network size is argued. It allows to extend the results to networks with arbitrary size and, thus, to real specimens.
\end{abstract}


\maketitle

\section{Introduction}
\label{Sec1}
The superconductive-normal state transition may occur according to diverse mechanisms, depending on physical conditions, material type and structure. In type II superconductors at $T\ll T_c$, where $T_c$ is the critical temperature in the absence of magnetic field and current, the transition occurs when fluxoids, injected  by external magnetic fields or strong bias current densities, begin to move causing energy losses and heating. This is relevant for the development of high-field superconducting magnets \cite{Heiden,Field,Marley,Chun,Togawa,Reichhardt,Lu,Kato,Das}. A different transition mechanism may occur when temperature is
close to $T_c$ at low current density. In this case, an
intermediate state  may be obtained, characterized by a mixture of superconductive and normal domains. This situation was first studied by Landau and Ginzburg in metals \cite{Landau,Ginzburg}. Recently, it has been considered to explain the excess noise in metallic or high-T$_c$ superconductors transition edge sensors (TES) used as bolometers to detect electromagnetic radiation at the level of
single photons \cite{Lindeman,Fraser,Brandt}.\\
The excess noise observed during a transition sheds light
on the microscopic processes underlying the transition itself
\cite{Carbone,Jobaud,Mazzetti08,Bid}.
In \cite{Mazzetti08}, the noise observed during the superconductor-normal transition in  granular MgB$_2$ films  has been ascribed to the subsequent formation of  resistive layers, with grains in the normal or in the intermediate state, between equipotential superconducting domains. The excess noise derives from the fact that each elementary event -the formation of a layer- implies the simultaneous resistive transition of several grains and, thus, gives rise to a voltage pulse  of rather high amplitude (\emph{avalanche noise}).\\
The present work is addressed to simulate the transition events occurring at granular level responsible for the avalanche-type noise in YBCO-like and MgB$_2$-like superconductors \cite{Mazzetti08,Mazzetti02}. The superconducting material is modeled as a network of nonlinear resistors, having Josephson junction current-voltage ($I-V$) characteristics with gaussian
distribution of critical currents.  The nonlinear resistors represent either weak links between grains (YBCO-like) or  grains  with strong links (MgB$_2$-like). In the strong link case, a couple or triple of resistors is used to represent two or three current components flowing through each grain respectively for two- (2D) and three-dimensional (3D) networks. The solutions of Kirchhoff equations for these networks are found by an iterative routine described in the next section. The main results of this analysis are: \begin{enumerate}
\item the resistive
    transition undergoes discrete step-like increments  both in weak and strong link materials.  The steps correspond to the creation of resistive layers constituted by grains or weak links in the normal or in the intermediate state.
    As temperature increases, grains or weak-links in the intermediate state gradually switch to the normal state. The trailing edge of the resistive transition grows more smoothly in MgB$_2$-like than in YBCO-like networks. This fact is related to the higher correlation when the elementary transition events occur in triplets rather than in independent nonlinear resistors.
\item the abrupt formation of resistive layers causes the large voltage noise observed at the transition in these materials. At the end of the transition, the resistive layers mix up. The resistance steps become smaller and the transition curve smoother. This smoothing results in noise suppression. A scaling law for the noise intensity is proposed in order to extend the results to larger networks, representing real materials. This effect was simply assumed in \cite{Mazzetti08}.  Here it is shown that the transition noise can be estimated once grain size and critical current distribution are defined.
\end{enumerate}

\section{Networks of Strong and Weak Links}
Before describing the details of the simulations, we provide a description of the main physical parameters  relevant to the electronic behavior of granular superconductors \cite{Beloborodov,Efetov}. In particular, it is worthy to remind that the phase transition boundary of granular superconductors is set by the value of the dimensionless tunneling conductance $g$:
\begin{equation}\label{g}
   g=\frac{G}{e^2/\hbar} \hspace{5pt},
\end{equation}
where $G$ is the average tunneling conductance between adjacent grains and ${e^2/\hbar}$ the quantum conductance.\\
Experiments show that samples with the normal state conductance greater than the quantum conductance (i.e. with the $g\gg 1$) become superconducting at low temperature \cite{Jaeger}, regardless of the ratio of Josephson  $J$ and Coulomb $E_c$ energies, defined respectively as:
\begin{equation}\label{J}
J=\frac{\pi}{2} g \Delta \hspace{5pt},
\end{equation}
with  $\Delta$ the superconductor gap, and
\begin{equation}\label{Ec}
E_c=\frac{e^2}{C_j}  \hspace{5pt},
\end{equation}
with  $C_j$ the grain capacitance.
This phenomenon can be accounted  by the electron tunneling between grains, in addition to the Josephson coupling \cite{Chakravarty}. The additional dissipative tunneling channel results in a reduction of the Coulomb energy to:
\begin{equation}\label{Ece}
   \widetilde{E_c}=\frac{\Delta}{2g}  \hspace{5pt},
\end{equation}
known as the effective Coulomb energy of the grain.
By comparing the Eqs.~(\ref{J}) and (\ref{Ece}), one can notice that for $g\gg 1$, $J$ is always larger than $\widetilde{E_c}$, implying a superconducting ground state, regardless of the Coulomb energy $E_c$. For $g\gg 1$, the granular superconductor can then be modeled within the mean-field BCS theory. Thus, its critical temperature is approximately  given by the single grain BCS critical temperature $2\Delta=3.53kT_c$. Conversely, for $g\ll 1$, the phase transition boundary between insulating and superconducting states is controlled by the ratio between $J$ and $E_c$. In this condition, by using a mean field approach, the critical temperature is given by $T_c=(1/4) z\pi g \Delta$, with $z$ the coordination number of the lattice \cite{Beloborodov,Efetov}.\\
The superconductor-normal transition in thin granular films with $g\gg 1$ can be modeled in terms of resistively shunted Josephson junctions, whose state is controlled only by the value of the normal resistance, rather than by  the Coulomb and Josephson energies.The simulations presented in this work  have been performed in the regime $g\gg 1$, to guarantee the onset of a superconductivity state at low temperature.\\
\begin{figure}
\includegraphics[width=7cm, angle=0]{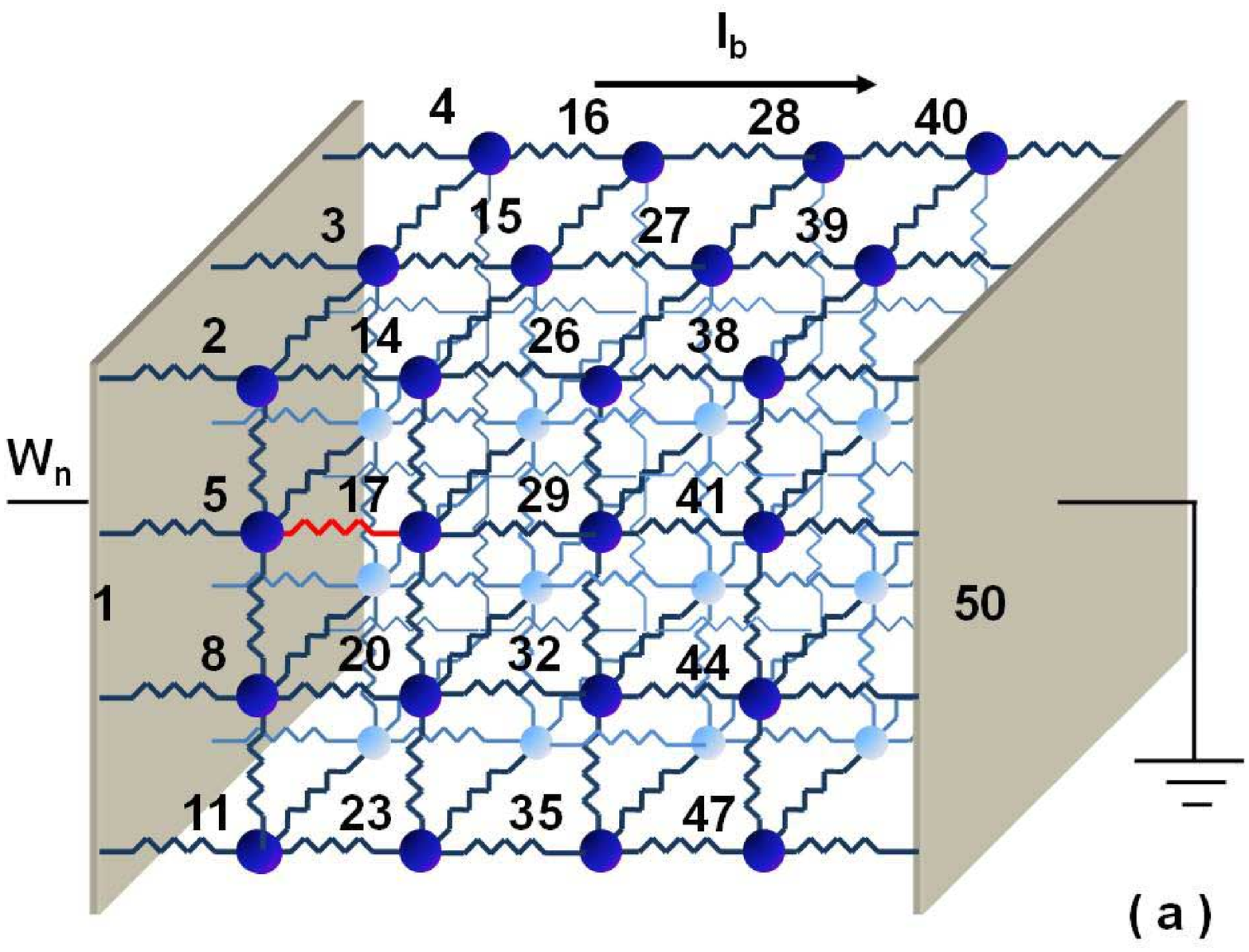}
\includegraphics[width=7cm, angle=0]{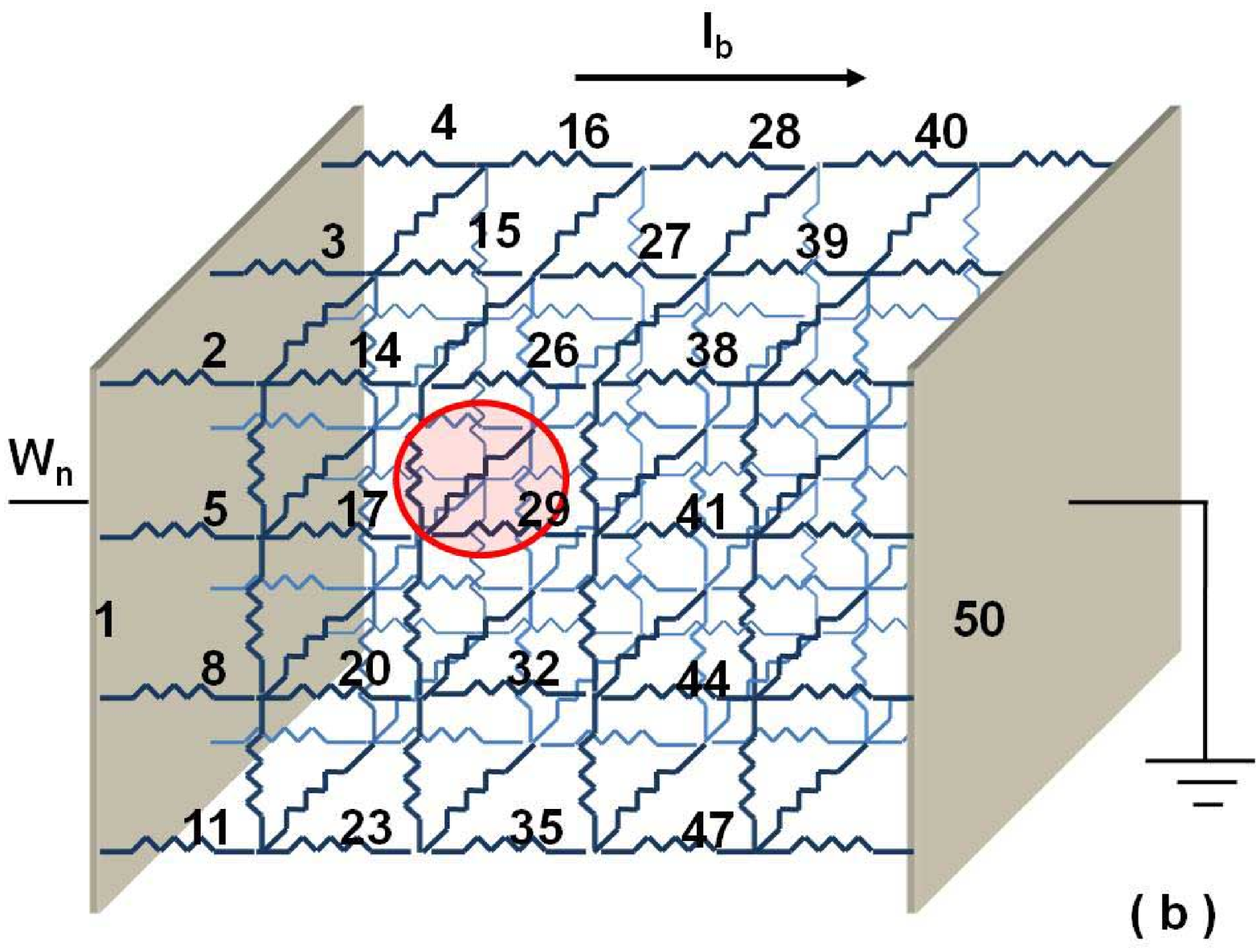}
\caption{\label{network}  Scheme of 3D networks of
granular superconductors  with: (a) weak-links
(YBCO-like) and (b) strong-links, where the transition involves directly the grains (MgB$_2$-like). The networks contain 48 grains. In (a), the grains are assumed to remain in the superconducting state during the transition, and correspond to the nodes of the network.  Each link between grains is a nonlinear resistor with the $I-V$ characteristic represented in Fig.\ref{IV}.In (b), the $I-V$ characteristic given in Fig.\ref{IV}  concerns the whole grain, which is represented by a triplet of orthogonal resistors (outlined by the red circle). The electrical conductance  of these resistors is determined from the $I-V$ characteristic of the grain by calculating the potential drop across the grain $i$ as $V_i=\left[\sum_{j=1}^3V_{ij}^2\right]^{1/2}$, where $j$ identifies the  three grains linked to the grain $i$. Since the grain is assumed to be isotropic, the conductances of the 3 resistors are assumed equal. The first (1) and last node (50) correspond to the electrodes.}
\end{figure}
\begin{figure}
\includegraphics[width=5cm, angle=0]{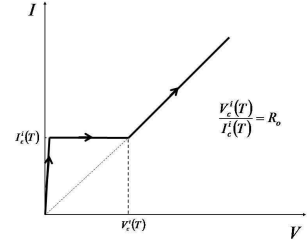}
\caption{\label{IV}  $I-V$ characteristics used for the simulation of the transition in the grain o weak-link networks. The resistance is assumed to be negligible ($10^{-10} \Omega$), when $I<I_c^i(T)$, while for $I>I_c^i(T)$ the normal state resistance is assumed to be $R_o$, equal for all the grains or links. The intermediate states occur for voltage drops between 0 and $V_c^i(T)=R_oI_c^i(T)$ and current $I=I_c^i(T)$. The critical current $I_c^i(T)$ is distributed according to a Gaussian distribution function for the different grains.
The transition from the superconductive to the intermediate state is set by the value of  $I_c$. The transition from the intermediate to the normal states is determined by the product  $V_c=R I_c$. Therefore the Gaussian distribution of the $I_c$ is enough to ensure the randomness of the   product $V_c=R I_c$ for all the grains. Furthermore, the quantity $V_c$ is directly related to the  Josephson time constant $\tau_J$, given by the Eqs.~(\ref{tauji}) and (\ref{taujr}),  that define the elementary switching time of the transition, and thus, are ultimately related to the behavior of noise.}
\end{figure}
In order to simulate the superconductor-normal transition in granular  materials, two types of  networks, shown in Figs.~\ref{network}(a) and (b), are considered. The networks are constituted by non-linear resistors, with Josephson junction characteristics, biased by a constant current generator. The resistive transition is estimated by solving a system of Kirchhoff equations, at varying temperature, for each network.\\
The network of Fig.~\ref{network}(a) refers to YBCO-like materials, characterized by weak-links \cite{Hilgenkamp}. In these materials, the transition occurs in two separated
steps: first, at lower temperatures, for the weak-links and, then, at higher temperatures,  for the grains. The network of Fig.~\ref{network}(a) is used to model the first stage of the transition, which involves only the weak-links, while
the grains remain superconductive.\\
The network of Fig.~\ref{network}(b) refers to MgB$_2$-like materials \cite{Larbalestier,Xi,Li}, whose transition involves directly the grains. Each triplet of resistors, outlined by the red circle, represents a grain. Since the current density within the grain may have any direction, the three resistors give a basis of three components of the current density for each grain. The current density sets the state (superconductive, intermediate, normal) of the grain according to its $I-V$ characteristic. For the sake of simplicity, all the grains are assumed to be isotropic and with the same average size, therefore the anisotropy effects are disregarded \cite{Lima,Sen,Choi}. This assumption is not limitative for what concerns the main aspects of the transition. It allows one to define a critical current $I_c^i(T)$ characterizing the grain $i$, according to a  Gaussian distribution, and a normal state resistance $R_o$ equal for all the grains. In real specimens, small changes of the grain stoichiometry influence the
critical current more than the normal state resistivity. The spread of the distribution of the critical currents and temperatures is responsible of the slope of transition curve \cite{Markiewicz}. The normal state  resistance $R_o$ is achieved when the current $I_i$ crossing the grain or the weak-link  exceeds $I_c^i(T)$.
The intermediate states are characterized by current $I_c^i(T)$ and voltage drop between $0$ and $V_c^i(T)$.
The $I-V$ characteristic of each non-linear resistor, representing a grain or a weak-link, is completely defined by the quantities: $I_c^i(T)$ and $R$.
The quantity $V_c=I_c^i(T) R$ is directly related to the  Josephson time constant by:
\begin{equation}
\label{tauji}
\tau_J=\frac{\Phi_o}{2\pi} \frac{1}{I_c R_{is}}
\end{equation}
for the intermediate states ($0<v<V_c$) and
\begin{equation}\label{taujr}
\tau_J=\frac{\Phi_o}{2\pi} \frac{1}{I_c R_{o}}
\end{equation}
for the normal states ($v>V_c$).
These time constants define the characteristic switching time during the transition, and thus, are ultimately related to the behavior of noise. In the next section, the resistive transition has been simulated in networks with (a) underdamped, (b) overdamped and (c) general $I-V$ characteristics, that are characterized by the Stewart-McCumber parameter $\beta_c=\tau_{RC}/\tau_{J}$, where $\tau_{RC}$ and $\tau_{J}$ are the capacitance and Josephson time constant. $\beta_c \gg 1$ in the case (a), $\beta_c\ll 1$ in the case (b) and $\beta_c\sim 1$ in the case (c). In particular, the onset of hysteresis has been analyzed upon cooling the granular system from the normal to the superconductive state.\\
When the transition involves the grains (strong-links), the current is given by:
$I_i=\left[\sum_{j=1}^3I_{ij}^2\right]^{1/2}$,
where $I_{ij}$ corresponds to the current flowing from the grain $i$ to its neighboring grains $j$ through each resistor of the triplet (see Fig.~\ref{network}(b)).
The $I-V$ characteristic is then used to find the value of the three resistors by means of an iterative routine to
solve Kirchhoff equations. The grains are assumed isotropic, thus the three resistors representing the grain will  have identical $I-V$ characteristics.\\
The simulations are carried on at constant bias current. The transition is caused  by the temperature increase, which reduces the critical currents of the grains or weak-links according to the following linearized equation:
\begin{equation}
\label{Ic_critical}
I_c^i(T)=I_{co}^i \left(1- \frac{T}{T_c} \right) \hspace{5pt},
\end{equation}
\noindent
where $I_{co}^i$ is the low-temperature critical current, distributed according to a  Gaussian function with standard deviation $\Delta I_{co}$ and mean value $I_{co}$.\\
The preliminary steps of the simulations are as follows:\\
\begin{enumerate}
  \item the  list of all the $N_o$  nodes of the network is created.
  \item the Gaussian distribution for the critical current $I_{co}^i$ is introduced. In Matlab, the vector $I_{co}^i$ is defined by:
      $I_{co}^i=I_{co} \left(1+ randn(N,1)\right)$,
      where $N$ is the number of junctions (individual resistors) for the network (a) or the number of grains (triplets of resistors) for the network (b). The quantity $randn(N,1)$  defines a set of $N$ random numbers extracted from a gaussian distribution having mean value  0 and variance 1.
\end{enumerate}
Then the iterative calculations are implemented as follows:
\begin{itemize}
  \item the vector $W_o$ of the tentative potential values is defined for all the $N_o$  nodes.
  \item for the network of Fig.\ref{network}(a), by using the $I-V$ characteristics, a conductance value $G_{ij}$ for each resistor between the nodes $i$ and $j$ is calculated
  \item else, for the network of Fig.\ref{network} (b), the conductance values, common to the three resistors
      representing each grain $i$, are calculated from the $I-V$ characteristics, by using the voltage drop:
      \begin{equation}
      \label{VoltageDrop}
      V_i=\left[\sum_{j=1}^3V_{ij}^2\right]^{1/2} \hspace{5pt}.
      \end{equation}
\end{itemize}
Once the $G_{ij}$ are known, the entries of the conductance matrix $\underline{\underline{G}}$ are:
\begin{subequations}
\begin{eqnarray}\label{G}
 G_{ij}&=& -G_{ij} \,\, (i,j =~\rm{contiguous}) \hspace{5pt}, \label{Ga} \\
 G_{ij}&=& 0 ~ \,\,\,\, (i,j =~\rm{not~contiguous})\hspace{5pt}, \label{Gb}\\
 G_{ii}&=& \sum_{k \epsilon V_i}G_{ik} \hspace{5pt}, \label{Gc}
\end{eqnarray}
\end{subequations}
where $G_{ik}$ are the conductances of the resistors connected to the node $i$.\\
Then, a new vector of node potentials $W_1$  is evaluated by solving the equation:
\begin{equation}
\label{Iinj}
\underline{\underline{G}} \cdot W_1 = I_{inj} \hspace{5pt},
\end{equation}
with respect to $W_1$.
$I_{inj}$ is a vector of dimension $N_o$, whose
elements are zero except the first one equal
to the bias current $I_b$. It represents the external current injected into node 1. The last node is grounded. \\
Then, the new set of  potentials $W_1$ allows to evaluate a new set of $G_{ij}$ and a new conductance matrix \underline{\underline{G}}. From Eq.~(\ref{Iinj}) an updated vector $W_2$ is then obtained. The iteration is repeated until the quantity $ \varepsilon={|W_n-W_{n-1}|}/{|W_n|}$ becomes smaller than a value $\varepsilon_{min}$ chosen to exit from the loop. In the present work, the simulations have been performed by varying $\varepsilon_{min}$ in the range $10^{-7} < \varepsilon_{min}< 10^{-11}$ to check that the value of $\varepsilon_{min}$ did not appreciably change the final solution. The total network resistance $R$ is then given by $W_n(1)/I_b$ for each value of $T/T_c$, where $W_n(1)$ is the potential drop at the contact ends.\\
The potential drops at the ends of each resistor for the case (a) and across the grain for the case (b) are compared to the values of the potential in the corresponding $I-V$ characteristics. Therefore, it is possible to distinguish weak-links or grains respectively in the superconducting, normal or intermediate state. Before discussing the simulations results, it is worthy to point to the different behavior of the two networks by introducing the intragrain conductance $g_{intr}$. For standard granular system, the condition $g \ll g_{intr}$ holds.\\
The intragrain conductance  of the weak-link network shown in Fig.~1(a) is much greater than 1 ($g_{intr}\gg 1$). The intragrain region is indeed assumed to remain in the superconducting state, since the transition occurs only at the weak-links. Conversely  for the strong-link network of  Fig.~1(b) the condition $g \sim g_{intr}$ holds, corresponding to an homogenously disordered granular system. This condition is consistent with the electronic properties of MgB$_2$-like superconductors \cite{Larbalestier}. The intragrain conductance $g_{intr}$ is related to the single grain Thouless energy $E_{Th}$ and to the interlevel spacing $\delta$ through
\begin{equation}\label{thouless}
    g_{intr}=E_{Th}/\delta \hspace{5pt}.
\end{equation}
When the energy $E_{Th}$ exceeds the mean level spacing $\delta$, it is $g_{intr}\gg 1$. The Thouless energy is defined by $E_{th}=D_o/a^2$, with $D_o$ and $a$ the diffusion coefficient and the radius of the grain. The interlevel spacing is defined  $\delta=1/(\nu V)$ with $\nu$ and $V$ the density of states at the Fermi energy and the volume of the grain.  The intragrain conductance strongly depends on the dirtiness of material and the radius of grain. These aspects are indeed relevant for MgB$_2$-like materials whose critical temperature is strongly dependent on  material quality,  atomic radii  and cell size \cite{Xi,Li}.

\section{Results}
Here, the successive stages of the resistive transition are simulated in granular superconducting materials either with  strong or with weak links. The superconducting material is represented as a network of nonlinear resistors having resistively and capacitively shunted Josephson junction characteristics \cite{book,Yu,Fazio}. We report the results of different simulations, carried on with 2D and 3D networks, both for grain and weak-link transition. In the simulations, the
transition occurs by increasing the temperature,  in proximity of the critical temperature $T_c$, starting from the superconductive state.

\subsection{Resistive layers in strong and weak-links networks}
\begin{figure}
\centering
\includegraphics[width=7cm,height=4cm,angle=0]{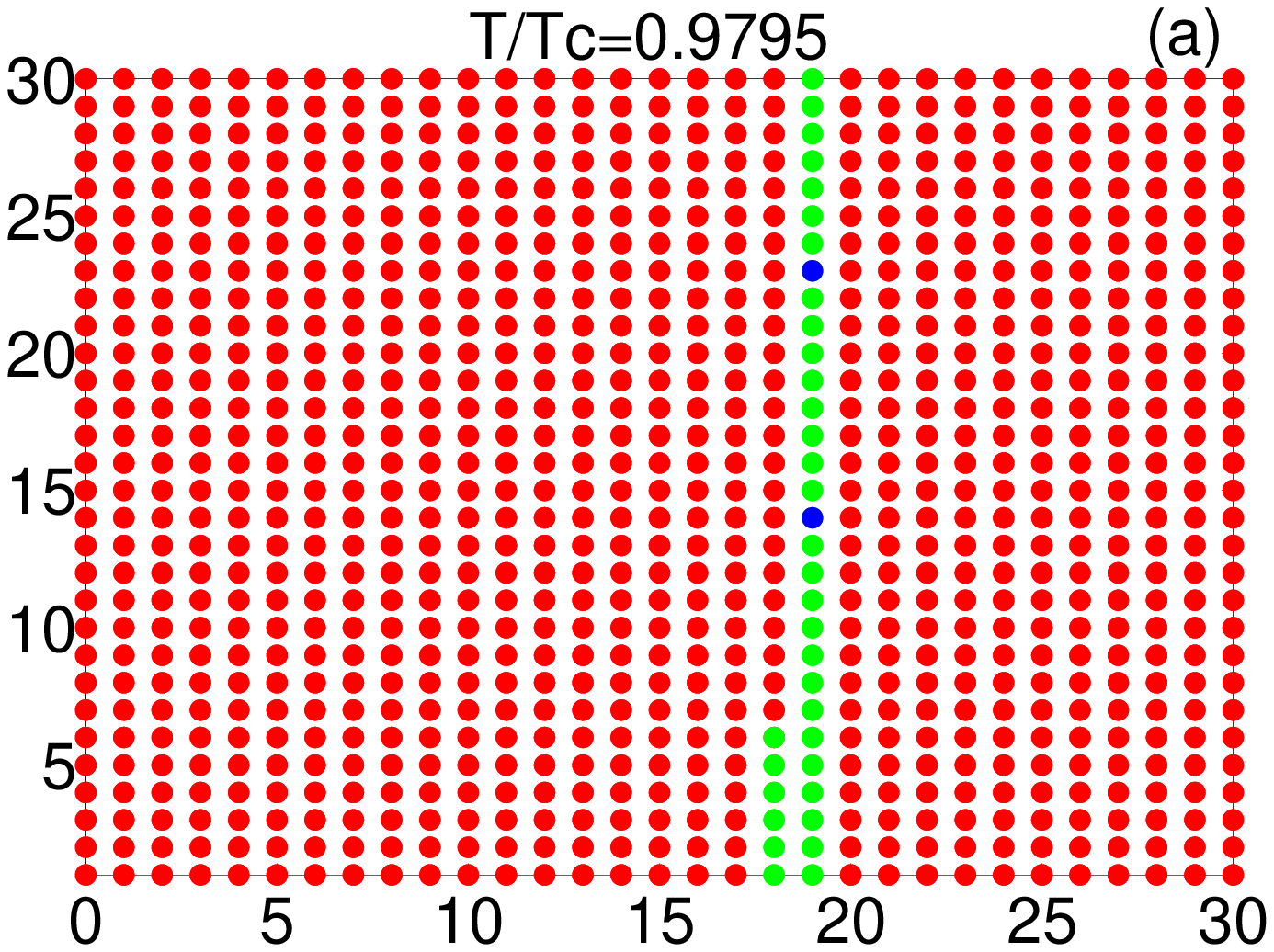}
\includegraphics[width=7cm,height=4cm,angle=0]{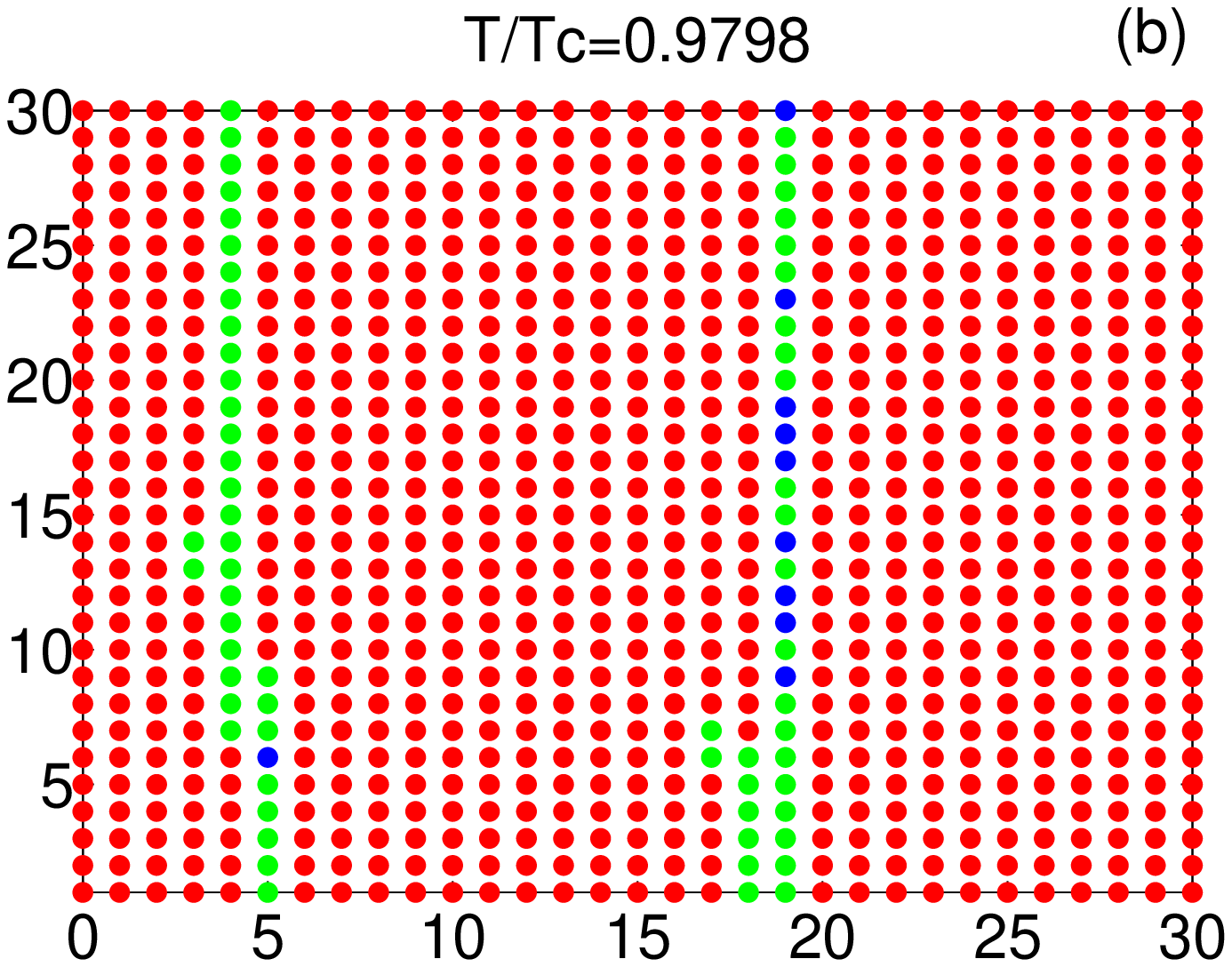}
\includegraphics[width=7cm,height=4cm,angle=0]{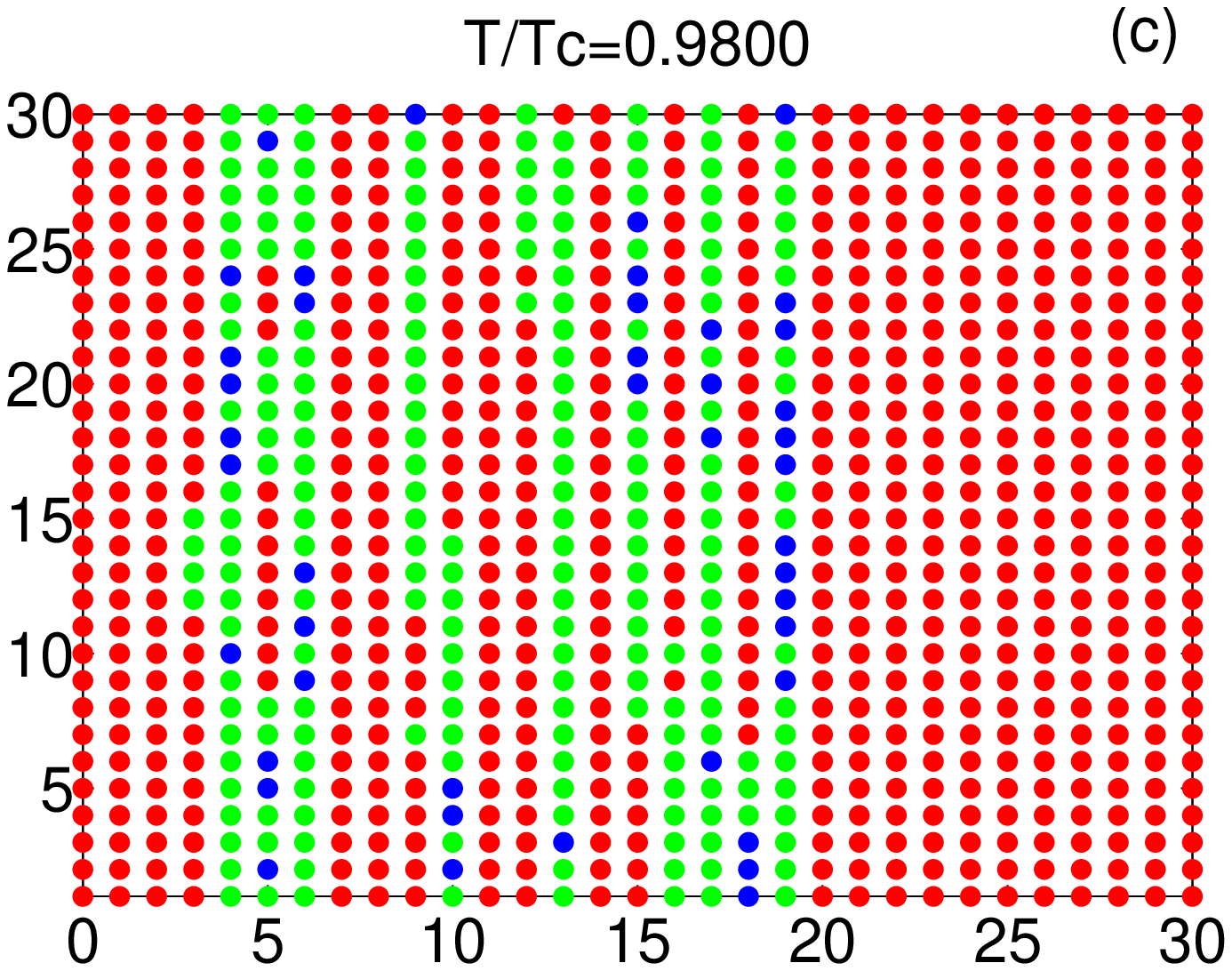}
\includegraphics[width=7cm,height=4cm,angle=0]{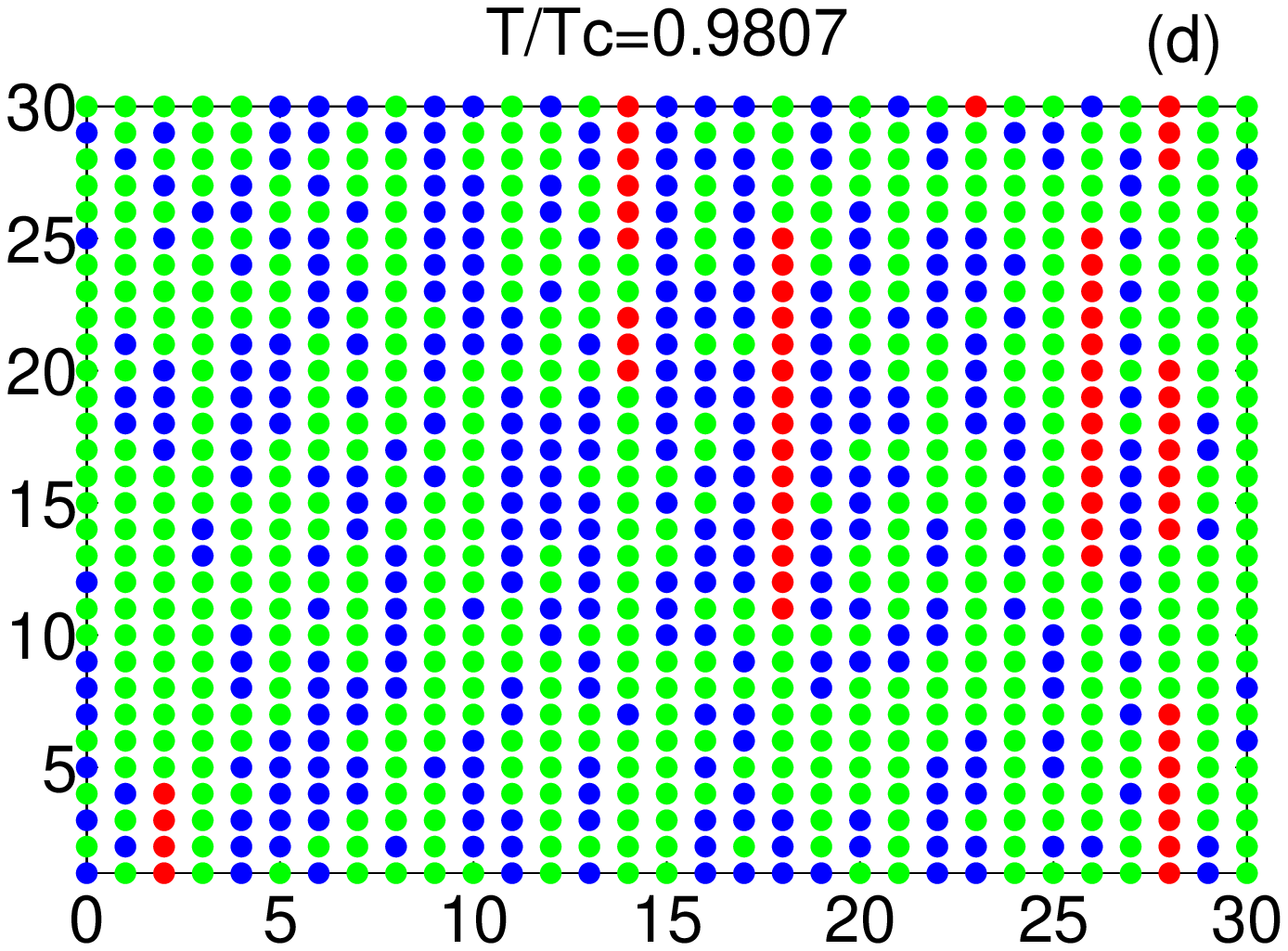}
\caption{\label{LayerGrain} Different stages of the superconductor-normal transition in a 2-dimensional MgB$_2$-like network with $30\times30$ grains. Red dots represent the superconductive grains, blue dots represent the
resistive grains,  green dots represent grains in the
intermediate state. In (a), the first resistive layer  (a strip in 2D) is formed, which corresponds to  the first step in the $R$ vs $T$ curve of Fig.~\ref{R2dgrain}. In (b), at slightly higher temperature the appearance of the second layer is shown, corresponding to the second step in Fig.~\ref{R2dgrain}~(a) etc. (c) The formation of more layers is shown. In (d), the situation at the transition end is shown. The layers  mix-up and the resistance steps become smoother, as shown in Fig.~\ref{R2dgrain}~(b). The parameters used in the simulation are : $T_c=39 K$; $R_o=0.32\Omega$; $I_b=1 mA$; $I_{co}=1.7mA$.}
\end{figure}

\begin{figure}
\includegraphics[width=7cm,angle=0]{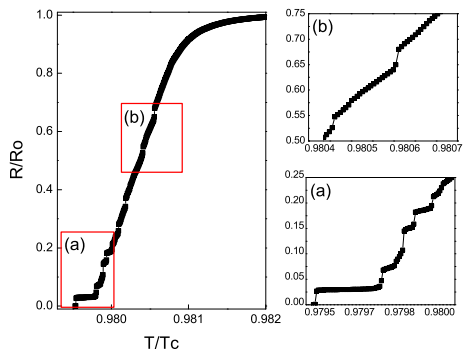}
\caption{\label{R2dgrain} Resistive transition  of
the two-dimensional network of $30\times30$ superconducting grains shown in Fig.~\ref{LayerGrain}. Figs.~\ref{R2dgrain} (a) and (b)  represent a zoom of the curve in two temperature intervals at the beginning and near the end of the transition. It may be noticed that the values of $T/T_c$ in correspondence of the steps in  Fig.~\ref{R2dgrain}~(a) correspond also to the layers represented in Figs. \ref{LayerGrain}~(a), \ref{LayerGrain}~(b), \ref{LayerGrain}~(c).}
\end{figure}

\begin{figure}
\centering
\includegraphics[width=7cm,height=4cm,angle=0]{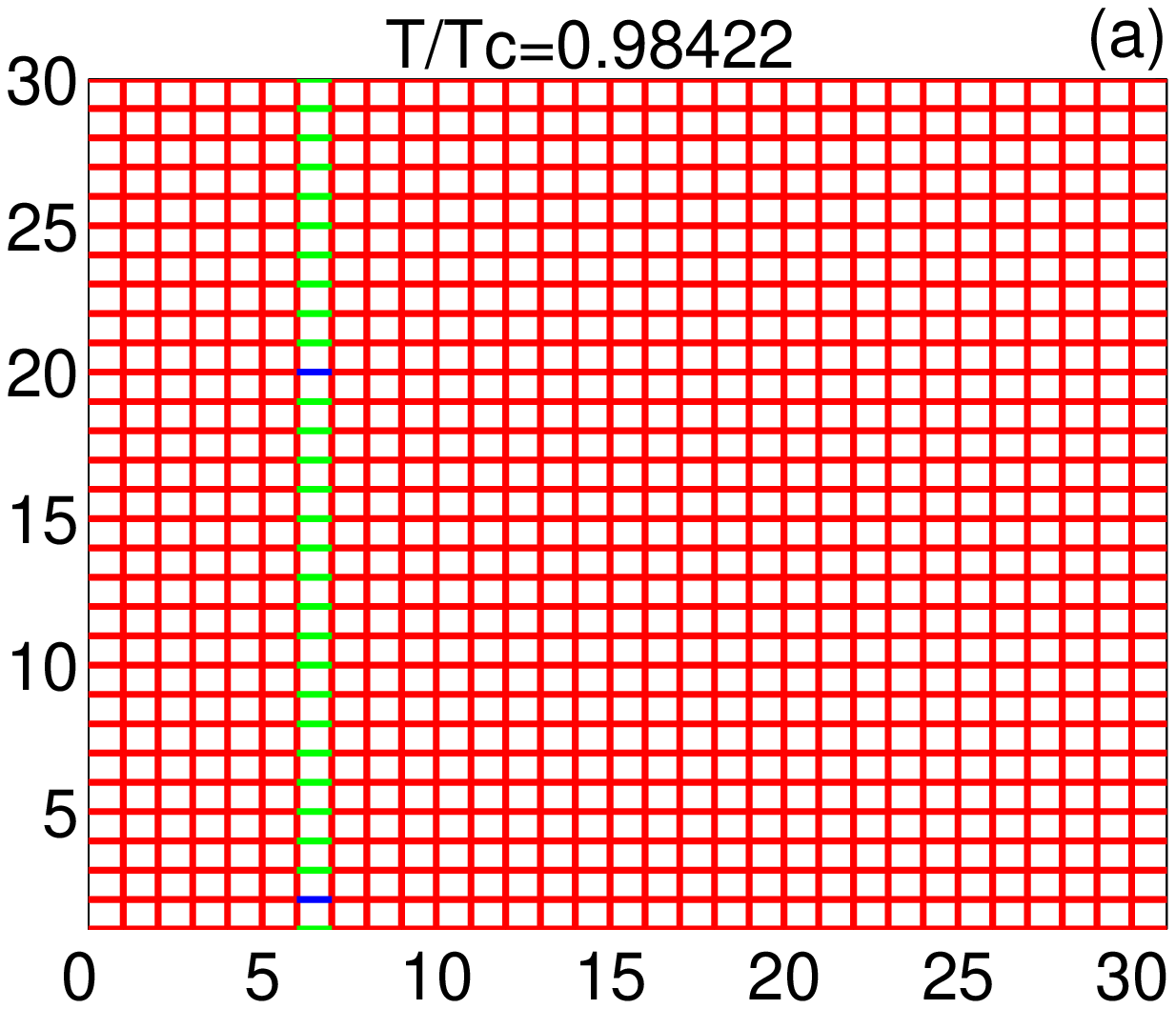}
\includegraphics[width=7cm,height=4cm,angle=0]{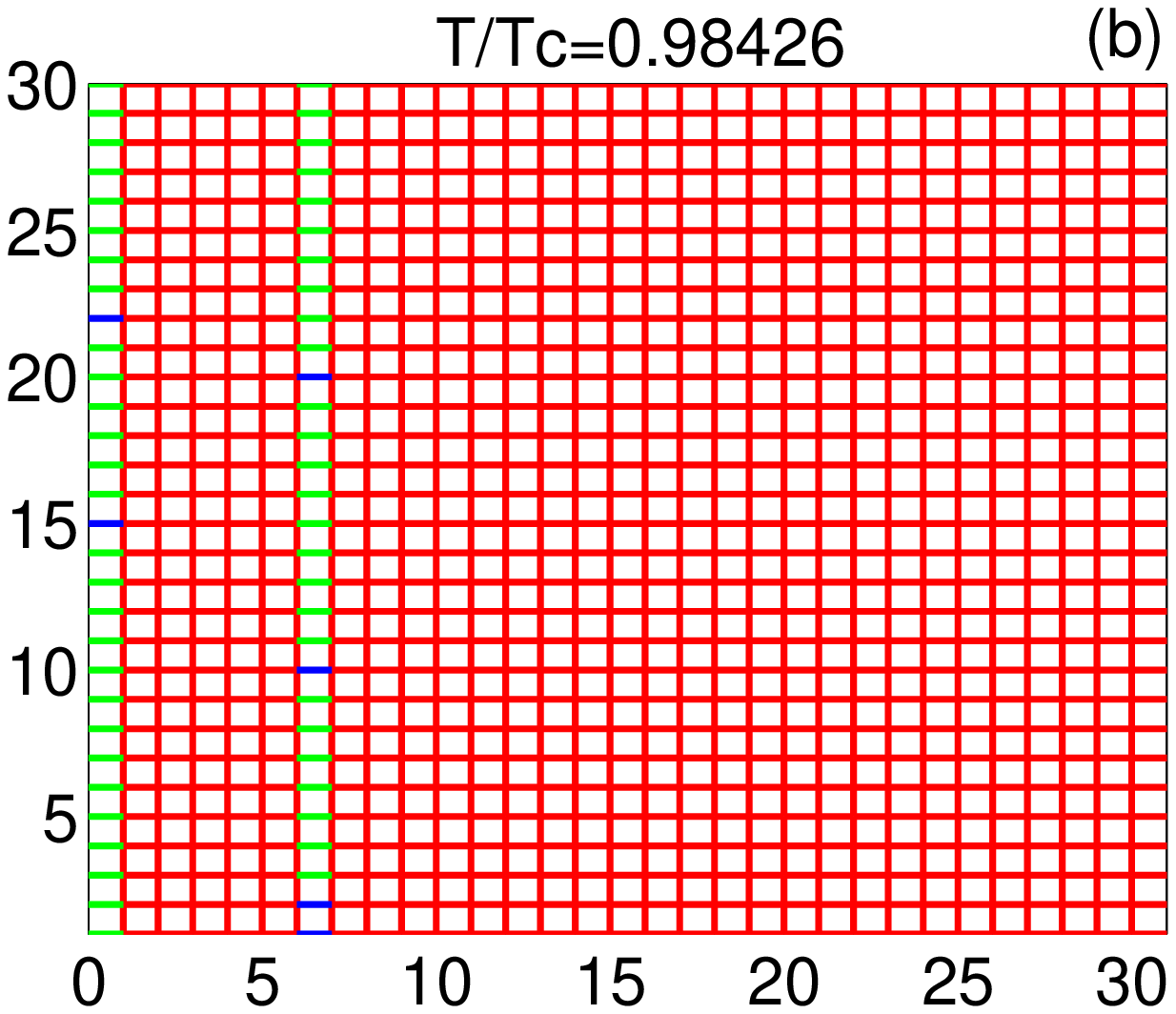}
\includegraphics[width=7cm,height=4cm,angle=0]{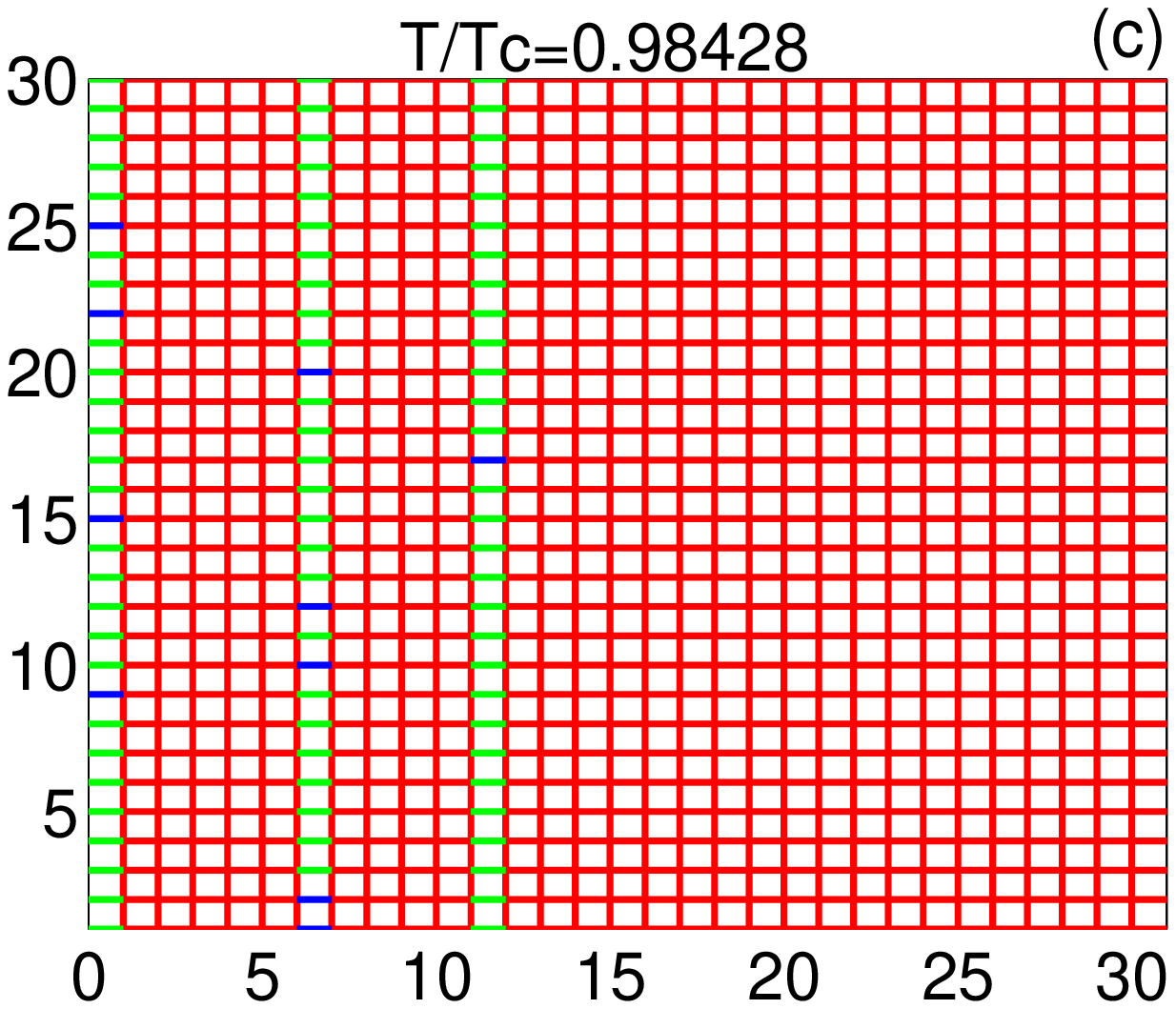}
\includegraphics[width=7cm,height=4cm,angle=0]{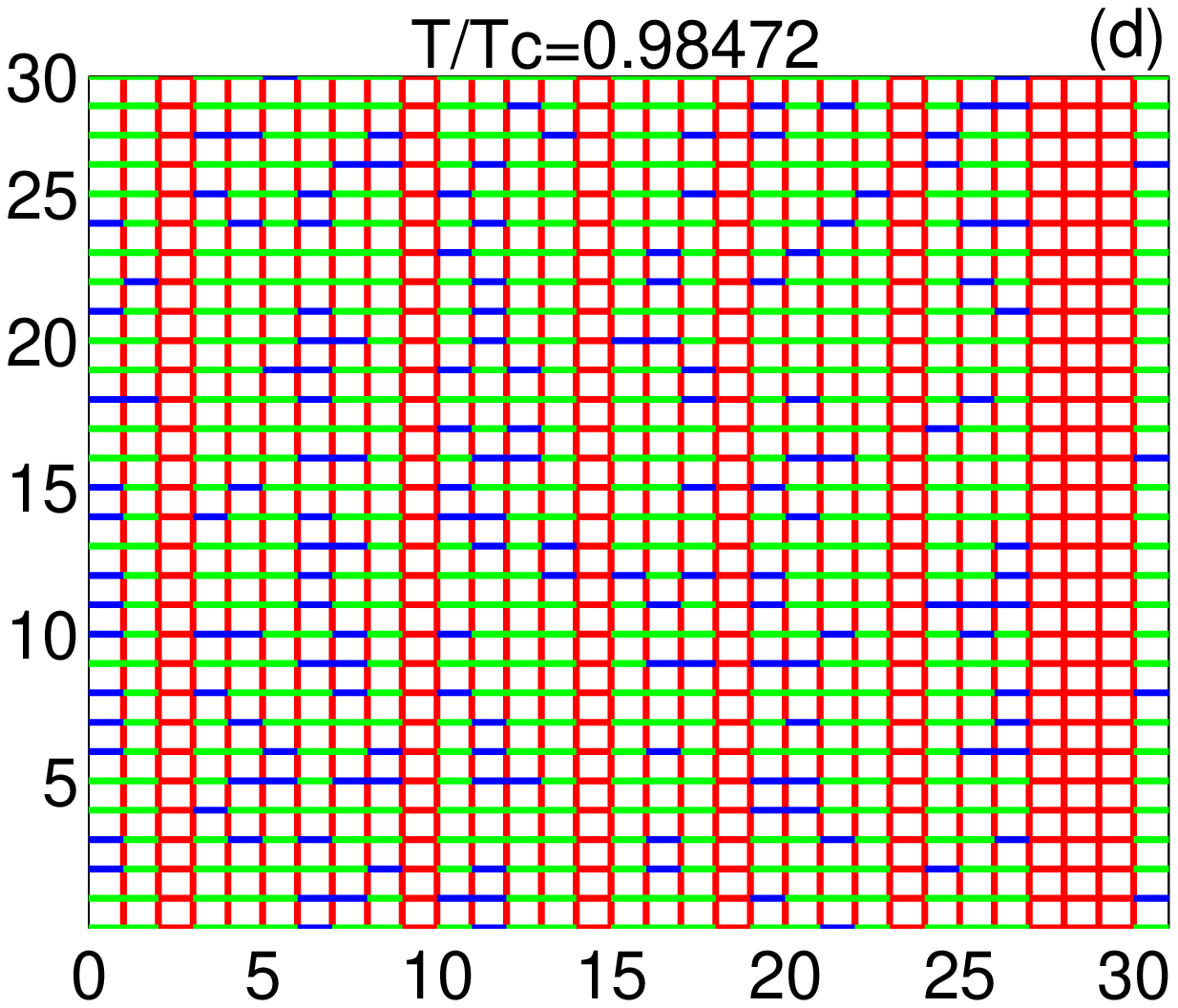}
\caption{\label{LayerLink}  Different stages of the superconductor-normal transition in a 2-dimensional  YBCO-like network  with $30\times30$ grains. Red lines represent
weak-links in the superconducting state, blue lines represent
resistive weak-links, while green lines represent weak-links in the intermediate state. (a) shows the formation
of the first resistive layer (a strip in 2D) and corresponds to the first step in the $R$ vs $T$ curve of Fig.~\ref{R2dlink}~(a). (b) and (c) show the transition at a temperature immediately following  the appearance of
the second and third layer and corresponds to the second and third step in Fig.~\ref{R2dlink}~(a) etc. (d) shows
the transition end, where the layers become mixed
up and the resistance steps become smaller, as shown in
Fig.~\ref{R2dlink}~(b). The parameters used in the calculations are: $T_c=65 K$; $R_o=0.32 \Omega$; $I_{b}=1 mA$; $I_{co}=2.1 mA$.}
\end{figure}

\begin{figure}
\includegraphics[width=7cm,angle=0]{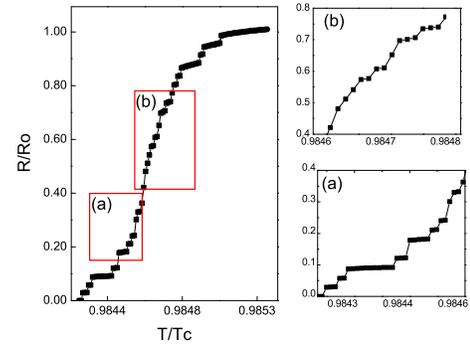}
\caption{\label{R2dlink} Same as Fig.~\ref {R2dgrain} but for the network of weak-links shown in Fig.~\ref{LayerLink}.}
\end{figure}

Figs.~\ref{LayerGrain} and~\ref{R2dgrain} refer to the resistive transition of a two-dimensional $30 \times 30$
network, representing a granular superconducting film of 900 grains characterized by strong links (MgB$_2$ type).\\
Figs.~\ref{LayerLink} and~\ref{R2dlink} refer to a two-dimensional $30 \times 30$ network, representing a  superconducting film of 900 grains characterized by weak-links (YBCO type).\\
$R$ and  $T$ are expressed as reduced quantities, namely as $R/R_o$, $T/T_c$. The relevant energy values and the parameters used for the simulations are reported in Table \ref{Table1}, in Table \ref{Table2} and/or in the figure captions.

\begin{table*}
\caption{\label{Table1} Relevant energy scales. The value of the capacitance used for the calculation of the Coulomb energy  are $2.7 \cdot 10^{-7}pF <C_j <2.7 \cdot 10^{-3}pF$.}
\begin{ruledtabular}
\begin{tabular}{cccc}
 &\multicolumn{1}{c}{$YBCO$}&\multicolumn{2}{c}{$MgB_2$}\\
 &  & $\pi$ bands
&$\sigma$ bands\\ \hline
  Superconductive gap  ($\Delta$) & 10 meV $\div$ 20 meV & 1.2meV $\div$ 3.7 meV & 6.4 $\div$ 7.2 meV \\
  Critical temperature  ($T_c$) & 65.8 K $\div$ 131.5 K & 7.9 K$\div$24.3K & 42.1 K$\div$47.4 K \\
  Ambegaokar-Baratoff product ($V_c={\pi \Delta}/(2e)$)& 15.7 mV $\div$ 31.4 mV & 1.9 mV$\div$ 5.8 mV & 10.1 mV $\div$ 11.3 mV\\
  Coulomb energy  ($E_c$)& 0.1 eV $\div$ $0.1 \cdot 10^{-4}$ eV & 0.1 eV $\div$ $0.1 \cdot 10^{-4}$ eV   & 0.1 eV $\div$ $0.1 \cdot10^{-4}$ eV \\
  Effective Coulomb energy  ($\widetilde{E}_c$) & $3.0$ $\mu$eV $\div$ $5.0$ $\mu$eV & $0.03$ $\mu$eV$\div$ $0.09$ $\mu$eV & $0.16 $ $\mu$eV $\div$ $0.18 $ $\mu$eV \\
  Josephson coupling energy ($J$)& 63.3 eV $\div$ 126.6 eV & 76.0 eV$\div$ 234.2 eV & 405.2 eV $\div$ 455.8 eV\\
\end{tabular}
\end{ruledtabular}
\end{table*}

\begin{table*}
\caption{\label{Table2} Simulation parameters.}
\begin{ruledtabular}
\begin{tabular}{ccccc}
 &\multicolumn{2}{c}{$YBCO$}&\multicolumn{2}{c}{$MgB_2$}\\
 & min& max& min
&max\\ \hline
  Low temperature Critical current ($I_{co}$) & 1.0 mA & 10 mA & 1.0 mA & 10 mA \\
  Normal state resistance ($R_o$)& 0.1 $\Omega$ & 3.0 $\Omega$ & 0.1 $ \Omega$ & $1.0 \Omega$ \\
  Critical voltage  ($V_c=R_oI_{co}$) & 0.1 mV & 10 mV & 0.1 mV &  10 mV \\
  Dimensionless tunneling conductance  ($g$)  & $1.3 \cdot 10^{5}$ & $4.3 \cdot 10^{3}$ & $1.3 \cdot 10^{5}$ & $1.3 \cdot 10^{4}$ \\
  Critical temperature  ($T_c$) & 65.8 K & 131.5 K & 11.8 K & 44.7 K \\
  \end{tabular}
\end{ruledtabular}
\end{table*}

\begin{figure}
\includegraphics[width=7cm,angle=0]{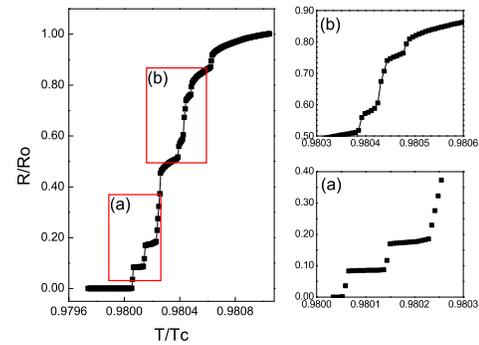}
\caption{\label{R3dgrain} Resistive transition  of a
three-dimensional network of $10\times 10\times 10$ grains. The parameters used in this simulation are  $R_o=0.32 \Omega$ and $I_{co}=1.7 mA$. Again each step should correspond to the creation of a layer of grains either in the normal
or in the intermediate state through the network cross-section.}
\end{figure}

\begin{figure}
\includegraphics[width=7cm,angle=0]{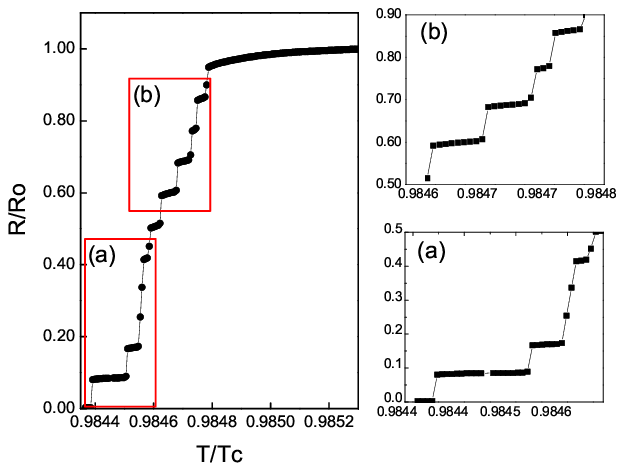}
\caption{\label{R3dlink}  Same as Fig. \ref
{R3dgrain} but for weak-links with the parameters: $R_o=0.32 \Omega$ and $I_{co}=1.7 mA$.}
\end{figure}
In Fig.~\ref{R2dgrain}, at the beginning of the
transition, the network resistance is zero, since all the grains are in the superconductive state. By effect of the temperature increase, a layer of grains either in the normal resistive (blue) or in the intermediate (green) state, crossing the whole film, is generated (Fig.~\ref{LayerGrain}~(a)). This layer must separate two equipotential superconductive regions  and, thus, the potential drop must be constant along the layer. Since the grains have different critical currents, the layer starts to form when the sum of  critical currents of its
grains  equals the bias current. The grain (or the weak-link) with the lowest critical current becomes resistive and set the voltage drop of the other grains in the layer. As temperature increases and critical current decreases, more and more grains in the intermediate state gradually switch  to the resistive states and the layer resistance increases.\\
As shown in Fig.~\ref{LayerGrain}, a resistive layer contains
at least one resistive (blue) dot and many intermediate (green) dots. Superconductive (red) dots are excluded since they would constitute a short. The formation of a resistive layer corresponds to a step in the $R~vs~T$ curve, as it can be seen by comparing Figs.~\ref{LayerGrain}~and~\ref{R2dgrain}.  Upon further temperature increase, other layers are created until the whole film undergoes the transition to the normal state.\\
At the beginning of the transition the layers are well separated  and have a thickness of approximately one grain. Correspondingly, the resistance steps shown in Fig.~\ref{R2dgrain} obey, as a good approximation, to a scaling law ($R/R_o = 1/30$ in the present case). At the transition
end, there is an intricacy of different layers and the
resistance increases smoothly with the temperature.
\\
Figs.~\ref{LayerLink} and ~\ref{R2dlink} correspond to granular superconductors characterized by weak-links (YBCO-like). The simulation refers to the resistive transition of the weak-links. The grains, represented by the nodes of the network, remain in the superconductive state. Also in
this case the transition occurs through the formation of resistive layers corresponding to resistance steps in the $R$ vs $T$  curve. \\ Figs.~\ref{R3dgrain} and ~\ref{R3dlink} report simulations carried on $10 \times 10 \times 10$
3D networks, representing superconductor films of 1000 grains respectively with strong and weak links. The presence of about 10 steps is expected from the scaling law holding before mixing up of the layers ($R/R_o = 1/10$).

\subsection{Hysteresis effects}
\begin{figure}
\includegraphics[width=5cm, angle=0] {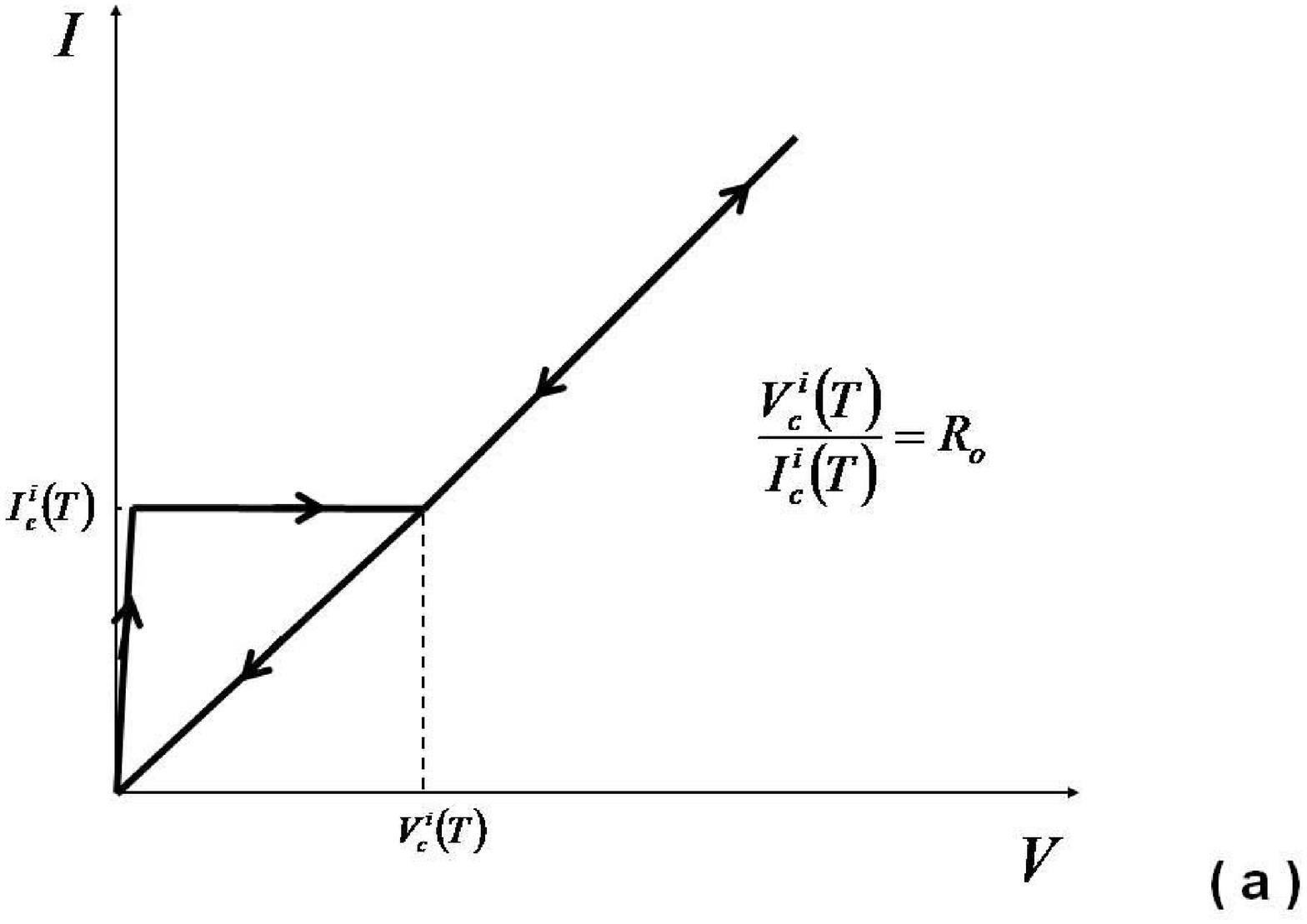}
\includegraphics[width=5cm, angle=0] {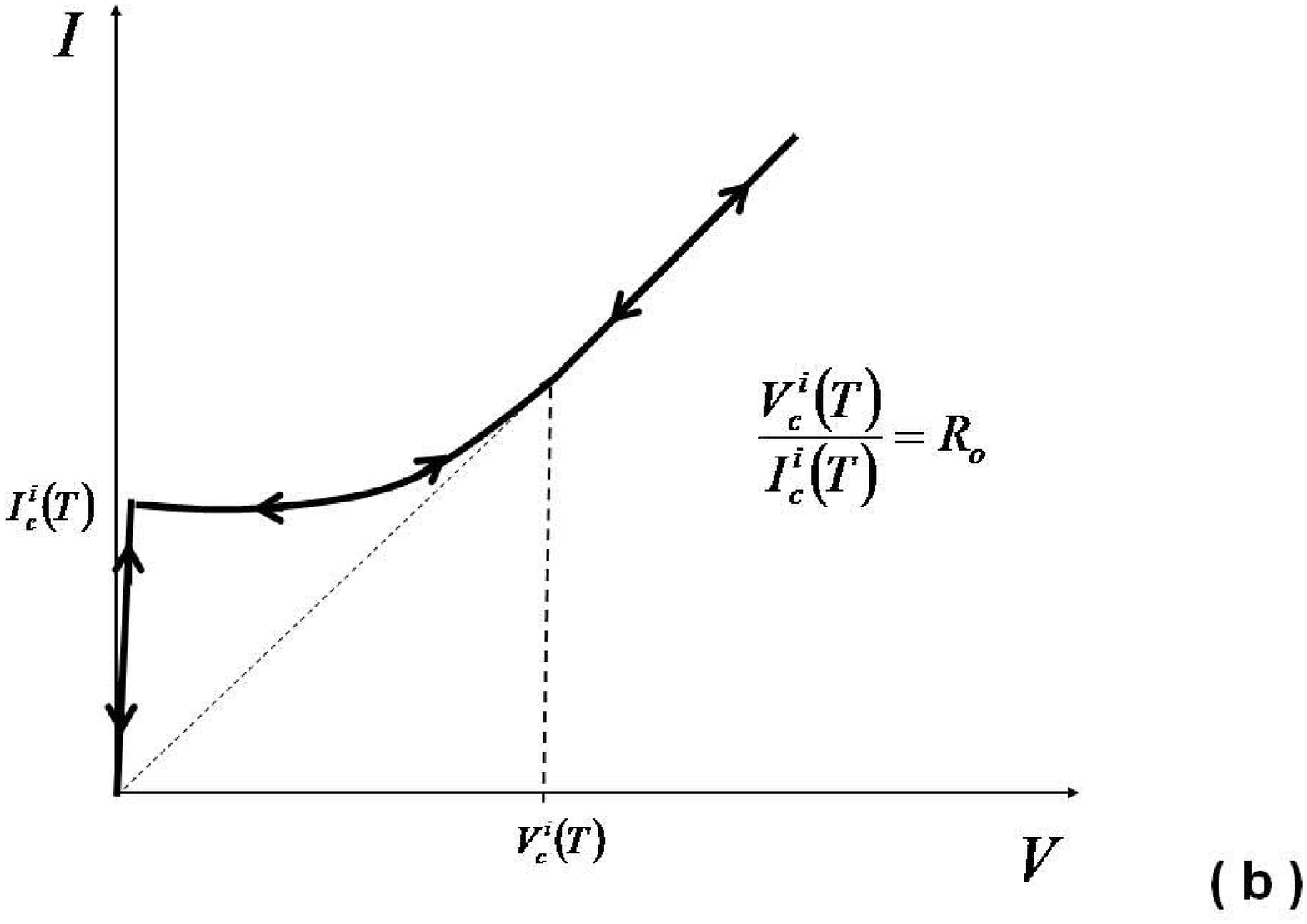}
\includegraphics[width=5cm, angle=0] {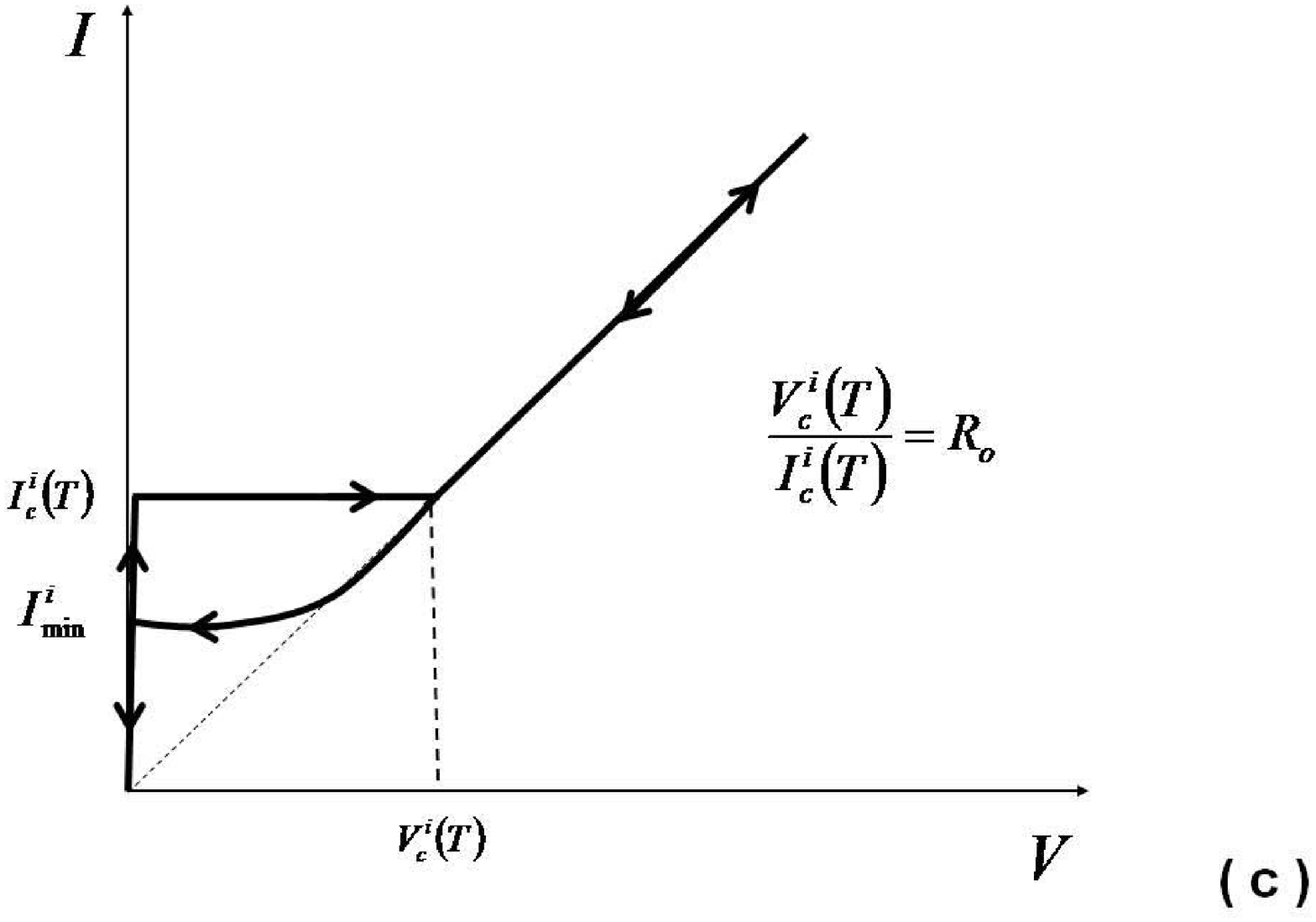}
\caption{\label{JJ} Josephson junction $I-V$ characteristics for grains or weak-links in case of underdamped (a), overdamped (b), generalized (c). $I_{min}^i$ depends on the Stewart-McCumber parameter $\beta_c$ and ranges from $I_c^i$ and 0 for $\beta_c \geq 0$, where $\beta_c=\tau_{RC}/\tau_{J}$, where $\tau_{RC}$ and $\tau_{J}$ are the capacitance and Josephson time constant respectively. $\beta_c \gg 1$ in the case (a), $\beta_c\ll1$ in the case (b) and $\beta_c\sim 1$ in the case (c)}
\end{figure}

\begin{figure}
\centering
\includegraphics[width=5cm, angle=0] {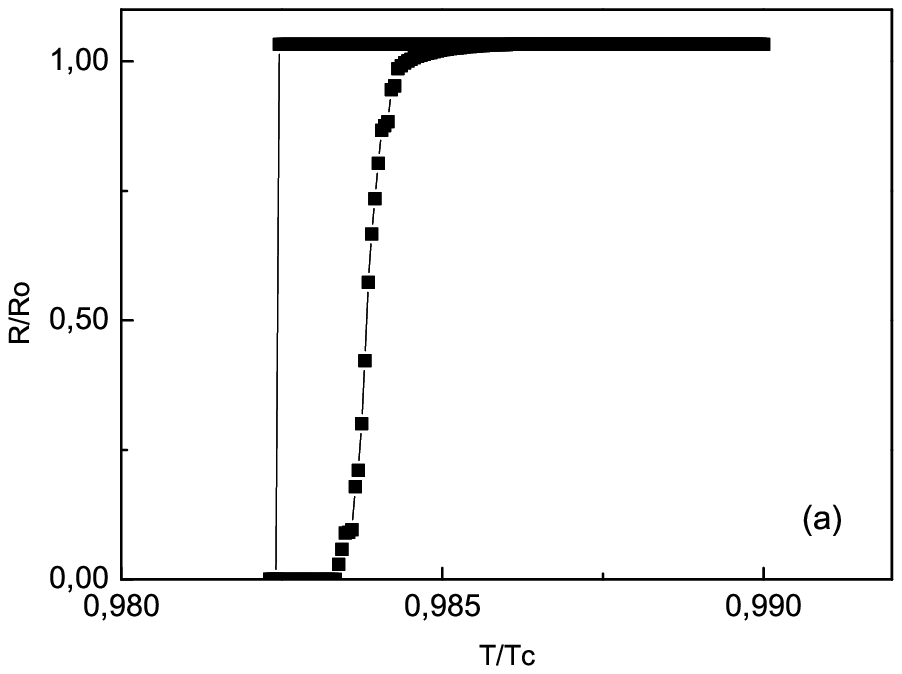}
\includegraphics[width=5cm, angle=0] {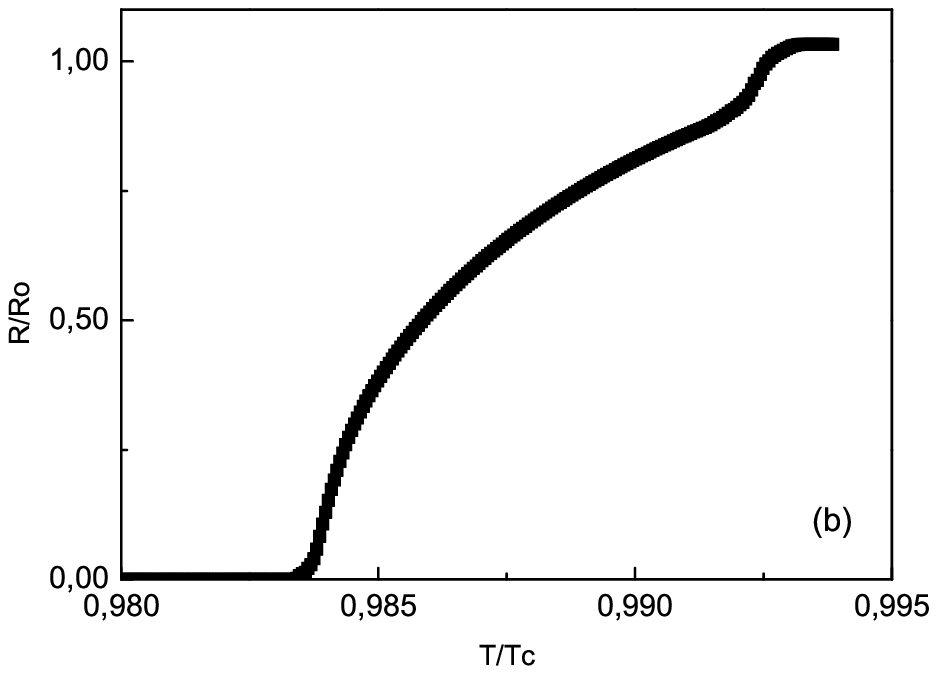}
\includegraphics[width=5cm, angle=0] {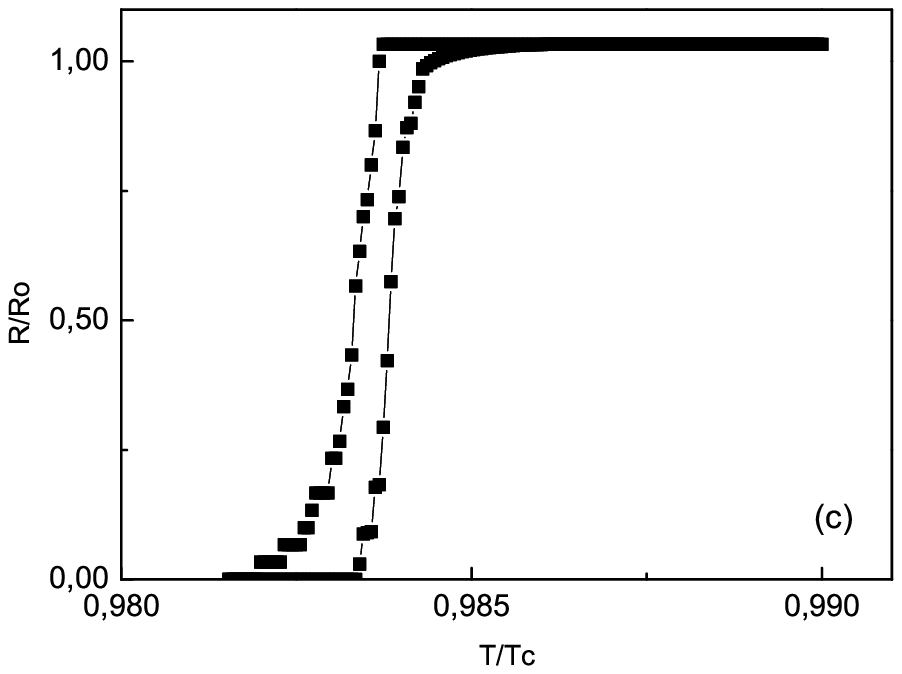}
\caption{\label{OLinkT}
Resistive transition  of the two-dimensional network of $30\times 30$ superconducting weak-links when temperature increases and decreases. The degree of disorder is 0.1 like the simulations.
The external current $I$ is kept constant at 1mA.
The Kirchhoff equation iteration is implemented with (a) underdamped Josephson junctions ($\beta_c \gg 1$), (b) overdamped Josephson junctions ($\beta_c\ll 1$) and general Josephson junctions ($\beta_c\sim1$).   One can note respectively hysteresis (a), no hysteresis (b) and partial hysteresis upon cooling the material from the normal to the superconductive state.}
\end{figure}

So far, we have been concerned with the superconductive-resistive transition as temperature increases with the main aim to investigate the successive formation of layers. Here, we investigate what the algorithm can predict when temperature is lowered and the superconductive final state is achieved starting from the normal one, thus addressing the hysteresis onset.
For this purpose, it is necessary to distinguish the $I-V$ characteristics of resistively shunted underdamped (a), overdamped (b) and generalized (c) Josephson junctions (shown respectively Figs.~\ref{JJ}(a),(b),(c)) \cite{book,Yu,Fazio}.
The curve $I-V$ (a) is hysteretic, the curve (b) shows no hysteresis, while the curve (c) exhibits partial hysteresis.\\
We have routinely solved the Kirchhoff equations of the strong and weak links networks by using the underdamped, overdamped and generalized $I-V$ characteristics and implementing a heating-cooling cycle around the critical temperature $T_c$.
For all the three cases: (\emph{i}) the conductance is $G=10^{10}S$  at  $I<I_c^i(T)$, (\emph{ii}) the normal state conductance $G_o=1/R_o$ at  $I>I_c^i(T)$ has been varied in the range reported in table \ref{Table2}, (\emph{iii}) $G$ and $G_o$ are much greater than the quantum conductance (i.e. $g\gg1$ always).\\ For the underdamped $I-V$ characteristics (a), the intermediate states are characterized by voltage drop in the range $0<V<V_c^i(T)$ and current equal to $I=I_c^i(T)$. The intermediate states correspond to the coexistence of superconducting and normal domains. Upon current (voltage) decrease starting from the normal state, the behavior is always normal resistive, implying that the system reaches the superconductive ground state without exploring intermediate states.\\
For the overdamped $I-V$ characteristics (b), the intermediate states are characterized by voltage drop in the range $0<V<2V_c^i(T)$ and current in the range $I_c^i(T)<I<I_c^i(2V_c^i(T))$, as described by the function:
\begin{equation}\label{overdamped}
    V=IR\sqrt{1-\left(\frac{I_c}{I}\right)^2} \hspace{5pt},
\end{equation}
instead of a constant value. The behavior of the overdamped Josephson junction is the same upon increasing and decreasing the current (voltage).\\
Fig.~\ref{JJ}(c) corresponds to the general case, the $I-V$ curve is partly hysteretic. Upon heating, the intermediate states are  characterized by a voltage drop in the range $0<V<2V_c^i(T)$ and current equal to $I_c^i(T)$. Conversely, upon cooling the intermediate states are described by Eq. (\ref{overdamped}).\\
Fig.~\ref{OLinkT} shows the resistive transition during a heating-cooling cycle in the case of a 2-dimensional network with weak links. In particular, Fig.~\ref{OLinkT} (a) refers to a network with underdamped weak links, where the maximum hysteresis effect can be observed. Fig.~\ref{OLinkT} (b) refers to a network with overdamped weak links, and no hysteresis is observed. Fig.~\ref{OLinkT} (c) refers to a network of weak links with generalized Josephson junction characteristic, where the amount of hysteresis is an average of the previous two cases.

\subsection{Avalanche Noise at the Transition}
The described results show that on increasing the temperature from the superconductive state, subsequent resistive layers are formed, until the whole specimen becomes normal. These layers are abruptly formed across the networks and correspond to step-like resistance increments in the $R$ vs $T$ curves. Each resistance step involves the simultaneous transition of all the grains in a layer, and corresponds to a voltage pulse at the end of the network, when a constant bias current $I_b$ is applied.  The number of these pulses during the transition is of the order of the number of grains or weak-links along the current direction. The large transition noise is due to the random superposition of these voltage pulses.
\\
For a given value of the normal state resistance, the step resistance  and the  step voltage amplitude are inversely
proportional to the number of steps. In real specimens with huge number of grains, the resistance steps cannot be resolved by static measurements of the transition curve. Conversely, the noise is a measure of the transition dynamics at granular level. The noise amplitude depends on the number and amplitude of discrete voltage steps and thus permits to justify the step-transition model. Assuming a Poisson distribution of the pulses, the power spectrum $\phi(\omega)$ of the noise is given by (Campbell theorem):
\begin{equation}\label{phi}
\phi(\omega)= \nu \overline{| S_v(\omega)|^2} \hspace{15pt},
\end{equation}
where $\nu$ represents the average number of pulses per unit time, and $\overline{| S_v(\omega)|^2}$
the average square modulus of the Fourier transform of each pulse.\\
Real materials correspond to very large networks, whose number of nodes is obtained by dividing the specimen dimensions by the average grain size.  The amplitude of the voltage pulses is inversely proportional to the number of steps, while $\nu$ is directly proportional to the number of steps along the transition curve. Since, in the low frequency limit ($\omega\rightarrow 0$), $\overline{|
S_v(0)|^2}$ is proportional to the square amplitude of the
voltage pulse, it turns out that the noise amplitude is inversely proportional to the number of steps. This shows that, for a given value of the network resistance, the voltage  noise amplitude in the limit of low-frequency is inversely proportional the number of grains along the direction of
the flowing current. This, in other words, means that superconductors with smaller grains are characterized by a lower intensity of the transition noise.

\section{Discussion and Conclusions}
The results reported above show several interesting aspects of the transition process in granular superconductors with weak and strong links.  One aspect, that can explain the
large noise observed during the resistive transition of
polycrystalline HTS, is  that the transition is not a
continuous dynamical process. The transition of a large
number of grains simultaneously occurs to form a resistive layer, approximately with the thickness of a single grain and orthogonal to the bias current density. This permits to evaluate the amplitude of the resistance steps generated by
the layer formation in real specimens on the basis of the average grain size and specimen dimensions. Moreover, a scaling law, deduced from the Campbell theorem, permits to deduce
the relation between the layer formation and the transition noise at low-frequencies.  The present approach gives exact numerical solutions for the transition. In addition it  allows to evidence the decrease of noise towards the transition end. By representing  the superconducting film as a
network of non linear resistors, it is possible to evaluate how the noisiness  decreases at the end of the resistive transition, according to the variance of the distribution of the grain or of the weak-links critical currents. This is a crucial issue for the development of superconductor based
sensors \cite{Sobolewski,Kreisler,Rahman}. The steepness of the $R$ vs $T$ curve gives higher photon detection signals (photoresponse) at the expenses of an increase of the noise.
Moreover, the resistance steps, corresponding to each layer formation, are visibly more squared and sharp for weak links than for strong-links transition. This fact may be related to the slope of the relative voltage noise spectra reported in
Refs.~\cite{Mazzetti08,Mazzetti02}. The power spectra are  $1/f^3$ and $1/f^2$-sloped in the range between a few Hz and 1kHz respectively for MgB$_2$ and YBCO. Since the power spectrum of a random staircase signal (i.e. a sequence of Poisson distributed exponential pulses, whose time decay tends to $\infty$) is exactly  $1/f^2$-sloped (i.e. a lorentzian function whose cut-off frequency tends to $0$), the rounding of the pulse trailing edge produces a steeper decay of the power spectrum, which tends to the $1/f^3$-slope. As a conclusion, it may be stated that the representation of granular superconductors as a network of nonlinear resistors with resistively shunted Josephson junction characteristics  add clues to the dynamics of the transition process. The assumption made in previous papers on the origin of the large transition noise in YBCO-like and MgB$_2$-like materials, is confirmed by the findings of the present work.


\end{document}